\documentclass[pdflatex, referee,sn-aps]{sn-jnl}% Chicago-based Humanities Reference Style
\usepackage{graphicx}%
\usepackage{multirow}%
\usepackage{amsmath,amssymb,amsfonts}%
\usepackage{amsthm}%
\usepackage{mathrsfs}%
\usepackage[title]{appendix}%
\usepackage{xcolor}%
\usepackage{textcomp}%
\usepackage{manyfoot}%
\usepackage{booktabs}%
\usepackage{algorithm}%
\usepackage{algorithmicx}%
\usepackage{algpseudocode}%
\usepackage{listings}%
\begin{document}

\title{Practical Classical Molecular Dynamics Simulations for Low-Temperature Plasma Processing: A review}

\author{\fnm{Pascal} \sur{Brault}}\email{pascal.brault@univ-orleans.fr}

\affil{\orgdiv{GREMI}, \orgname{CNRS - Universiy of Orl\'eans}, \orgaddress{\street{14, rue d'Issoudun BP6744}, \city{Orl\'eans Cedex 2}, \postcode{45067}, \country{France}}}


\abstract{Molecular Dynamics simulations are becoming a powerful tool for examining and predicting atomic and molecular processes in various environments. The present review showcases how Molecular Dynamics simulations can provide valuable insights into various processes in the fields of plasma physics, chemistry, and interactions with materials and liquids. Some notable processes include gas phase polymerization, deposition, plasma-catalysis, discharge breakdown, and vibrational excitation.}




\maketitle

\section{Introduction}\label{sec1}

Analyzing low-temperature plasma processes using Molecular Dynamics (MD) simulations began in the mid-1960s to mid-1970s \cite{Hockney1966, Birdsall1969, Hockney1973, Hockney1974}. However, it remained largely marginal, except in the field of dense (warm) plasma. Interest in using MD simulations in the field of low-temperature weakly ionized plasma emerged in the mid-1980s through the study of electron-ion recombination, utilizing MD and Monte-Carlo simulations \cite{Morgan1984}. The availability of 3- and many-body interatomic potentials, necessary for performing MD simulations, emerged between 1985 and 1990 \cite{Stillinger1985, Tersoff1988, Tersoff1988a, Tersoff1988b, Brenner1989, Brenner1990, Daw1983, Daw1984}, and this truly propelled the MD simulation approach to plasma etching and deposition. Since the publication of these original articles, many refinements have been made to these forcefields, which are still in use today by S. Sinnot et al. \cite{Sinnott2012}.
The first step in developing MD simulations for low-temperature plasma processes was atom deposition or fast atom deposition, mimicking low-energy sputtering deposition or ion beam deposition. In the latter case, ions are treated as fast neutrals, assuming rapid neutralization of the ion near the surface \cite{Muller1987b, Muller1987a, Muller1987}. These simulations were often performed in a two-dimensional framework due to computational performance limitations at that time.
With the advent of 3-body and many-body potentials \cite{Sinnott2012} and the availability of High-Performance Computers, plasma etching and deposition have been studied to compare and understand experimental results (e.g., pioneering works on chemical and physical sputtering of silicon \cite{Barone1995} and deposition of thin films \cite{Luedtke1989, Gilmore1991}).
During this period, only three dedicated reviews to Molecular Dynamics studies of low-temperature plasma processes were published \cite{Gou2008, Graves2009, Neyts2017}. The present review aims to provide an updated account of the progress made using this simulation technique and its application to plasma processing.  The next section (Section~\ref{sec2}) will address how handling MD simulations and how including experimental conditions for enabling comparison between simulations and experiments, while Section~\ref{sec3} will focus on the plasma processes that can be described by MD simulations.
 
\section{Practical MD simulations}\label{sec2}
\subsection{Principles}\label{subsec21}
Basically, MD simulations involve solving the Newton equations of motion for a collection of atoms, molecules, or particles \cite{haile1992, frenkel2002, rapaport2004, Allen2017}. For a system with $N$ atoms, with coordinates $\{\vec{r}_i \}_{i=1,...,N}$, an interaction potential $V = V(\vec{r}_1, \vec{r}_2, ...,\vec{r}_n)$, and $m_i$ representing the mass of atom $i$, the Newton equation at time $t$ can be expressed as\\
\begin{equation}
\label{eq1}
 m_i \frac{d^{2} \vec{r}_i(t)}{d t^{2}} = \vec{f}_i(t), \quad \mbox{with} \quad
 \vec{f}_i(t) =  -\frac{\partial V(\vec{r}_1(t), \vec{r}_2(t), ...,\vec{r}_n(t))}{\partial \vec{r}_i(t)} 
\end{equation}

where  $ \vec{f}_i(t)$ is the force acting on atom $i$.
Moreover, solving Equation~(\ref{eq1}) involves determining the trajectories of each individual species in the system being considered. MD can be used to treat systems with a large number of species, up to 10$^9$, utilizing high-performance computing facilities. For solid or liquid systems, this corresponds to nanoscale sizes, where 1 $\mu$m of solid or liquid matter contains around 10$^{11}$ atoms.
Solving Equation(\ref{eq1}) requires knowledge of the interaction potentials and a set of two initial conditions: positions and velocities of all species. The availability of robust semi-empirical interaction potentials allows for simulations to be run in a reasonable amount of computer time for a given system size and complexity. If these potentials are not available, the interaction potential can be evaluated using quantum chemistry methods during the time integration of the Newton equations of motion. Since the pioneering work by Car and Parrinello \cite{Car1985}, many improvements have been made to popularize Ab-Initio Molecular Dynamics (AIMD, also called First-Principles MD or FPMD) \cite{AIMD2004}. The main focus has been on developing fast and accurate algorithms to overcome the inherent computational cost of these methods \cite{Grimme2017}. In recent years, Machine Learning approaches have also been developed to provide accurate interaction potentials, leading to faster algorithms than AIMD and being "universal", such as the very recent M3GNET 3-body potential parametrization \cite{Chen2022}. Section\ref{subsec24} provides a description of the most commonly used interaction potentials. The terms "interaction potentials", "interatomic potentials", and "force fields" will have the same meaning throughout this review.
In addition to interaction potentials, the initial conditions are also important for connecting MD simulations to the real world of experiments. The simplest way is to select initial positions, which can correspond to well-defined positions such as crystalline states or be randomly selected to represent amorphous, liquid, or gas phases. For initial velocities, the best approach is to randomly select them from "real" distributions, such as Maxwell-Boltzmann for thermalized processes or sputtering distributions for sputter deposition \cite{Xie2014}. Alternatively, experimental energy distributions obtained through energy-resolved mass spectrometry or laser diagnostics can be used.
Another important consideration is the "small" size of MD simulation boxes, which does not allow for excessive energy dissipation, for example, from bond formation or energy transfer to the surface of materials or liquids. There are not enough species in the system to share the excess energy and maintain the temperature throughout the simulations. The solution is to use thermostats that are set up to absorb this excess energy in a manner consistent with the studied process. Since thermostats operate over a damping time, it is important to choose a damping time that is consistent with the realistic energy dissipation time \cite{Hou2000} observed in experiments \cite{Pentecoste2016}.

\subsection{Which species, which phenomena ?}\label{subsec22}
\ref{tab1} displays all plasma species that can be treated using MD simulations \cite{Neyts2017}. It should be pointed out that including electrons in a MD simulation is a problem since their mass is $m_e = 5.49 \times 10^{-4}$ amu. Therefore, sampling the interaction potential will require an integration timestep of the equations of motion in the subattosecond range, instead of the femtosecond range for heavy species. Consequently, including electrons will dramatically increase the computational time. Fortunately, to date, there exist two force fields that explicitly include electrons: eFF \cite{Su2007, J-Botero2011} and e-reaxFF \cite{Islam2016, Leven2021, Akbarian2021}. These force fields overcome this limitation by setting the electron mass to 1 amu, effectively considering the electron as an unreactive negative hydrogen ion. The eFF force field is primarily used for describing warm dense plasmas, while e-reaxFF is employed for electron charge transfers in condensed matter and molecules \cite{Islam2016a, Islam2016}, and more recently for electrical breakdown \cite{Akbarian2021}.
Diffusion in atmospheric plasma has been addressed using Molecular Dynamics simulations, incorporating ion-ion strong coupling \cite{Acciarri2022, Acciarri2023}.

\begin{table}[h]
\caption{Possible included plasma species in MD simulations and associated processes.}\label{tab1}%
\begin{tabular}{@{}lll@{}}
\toprule
Plasma component & inclusion possible ?  & Addressed phenomena \\
\midrule
atom and hyperthermal specie & yes  & deposition, etching\\
molecule and radical  & yes & plasma chemistry, etching, deposition  \\
ions   & yes & sputtering, reactivity \\
electron   & yes & Electrical breakdown, e- attachment \\
photon & yes   & polymer degradation, laser sputtering\\
electric field    & yes & e-field assisted processes\\
electronically excited states    & yes   & etching\\
vibrationally excited states   & yes  & plasma catalysis, dissociative chemisorption\\
\botrule
\end{tabular}
\end{table}

\subsection{Relevance for comparison to experiments}\label{subsec23}
The question of comparison with experiments is closely related to which statistical information can be recovered with such small simulation boxes: for comparison, a plasma reactor can range in size from micrometers (micro-plasma) to meters (large plasma materials treatment facilities). For solid and liquid states, there is almost no problem in deriving statistical information (such as diffusion coefficients) since the density is on the order of a few tens of species per nm$^3$, resulting in more than 10,000 species per 10x10x10 nm$^3$ box volume. However, for gas phase species, the situation is less favorable since at a pressure of 10$^5$ Pa (1 atmosphere), the gas density is only 2.4 x 10$^{-2}$ nm$^{-3}$. This requires a large box size. To reach around 10,000 atoms, a volume of 75x75x75 nm$^3$ is required. It should be noted that even when using a large interaction potential cutoff length of 2 nm, the simulations will calculate straight trajectories without any interactions for more than 90% of the computer time. Nevertheless, such simulations can still be carried out \cite{SwagatikaMishra2023}.
When dealing with lower pressures, such as those encountered in low-pressure plasma processing, it becomes necessary to reduce the box size while maintaining the correct description of interactions. To achieve this, we can consider that the collision number for a given experiment $n_{\rm exp}$ should be the same in the corresponding MD simulation $n_{\rm sim}$. In this case, equating the collision numbers in both situations gives the scaling relation \cite{brault2018}:
\begin{equation}
\label{eq2}
P_{\rm sim}.d_{\rm sim} =  P_{\rm exp}.d_{\rm exp}
\end{equation}
$d_{\rm  exp,sim}$ being the typical experiment and simulation box dimensions, respectively. From Eq.~(\ref{eq2}) the number of species $N_{\rm sim}$ in the simulation box can be deduced, using $V_{\rm sim} =d_{\rm sim}.S_{\rm sim}$, $V_{sim}$ and $S_{\rm sim}$ being the simulation box volume  and  basal surface area:
\begin{equation}
\label{eq3}
N_{sim} =  \frac{P_{\rm exp}}{k_B.T_g}S_{\rm sim}d_{\rm exp}
\end{equation}
with $T_g$ being the temperature of the plasma neutral and ion species assuming perfect gas theory. Since the goal is to limit the computer time used for calculating straight trajectories without interaction, it is sufficient to have the distance $l = \left(V_{\rm sim}/N_{\rm sim}\right)^{1/3}$ between species in the simulation box, just greater than the largest interaction cutoff distance $r_c$ of the system. Which leads to:
\begin{equation}
\label{eq4}
d_{sim} > \frac{N_{\rm sim}}{S_{\rm sim}}r_c^3
\end{equation}
which reduces to $d_{\rm sim} > N_{\rm sim}^{1/3} r_c$ for $S_{sim} = d_{sim}^2$.
For Coulomb interactions, the cutoff value is the largest short-range cutoff length, when k-space integration is used for the long-range part of the interaction potential. 
Since low-pressure plasmas are often used for deposition/etching processes, a link between simulated $\tau_{sim}$ and experimental $\tau_{\rm exp}$  etching/deposition rates can be drawn. In a first attempt, it is enough to consider that sticking coefficients in experiments $\sigma_{\rm exp}$ and MD simulations $\sigma_{\rm sim}$ should be the same (equivalently to collision numbers in gas phase). Thus, the experimental rate $\tau_{\rm exp}$  can be predicted as:
\begin{equation}
\label{eq5}
\tau_{exp} = \sigma_{\rm sim}.\phi_{\rm exp}
\end{equation}
where $\phi_{\rm exp}$ is the experimental flux of ions of neutrals to the surface.

\subsection{Interaction Potentials}\label{subsec24}
Many interaction potentials are suitable and employed in molecular dynamics simulations. The most simple ones are pair potential like Lennard-Jones and Morse potentials \cite{Rappe1992, Jacobson2022}. Due to their pair nature, they allow very fast calculations on large systems. But for detailed calculations, there is a need for more accurate interatomic potentials.

For metal atoms, the Embedded Atom Method \cite{Daw1983, Daw1984, Johnson1989, Daw1993, Zhou2001, Zhou2004} is a popular many-body forcefield which use the concept of electron (charge) density to describe metallic bonding. Thus, the energy of a solid is a unique functional of the electron density, for which each atom contributes through a spherical, exponentially-decaying field of electron charge, centred at its nucleus, to the overall charge density of the system. Binding of atoms is modelled as embedding these atoms in this “pool” of charge, where the energy gained by embedding an atom, at location r, is some function of the local density. In this frame, the total energy reads:
\begin{equation}
\label{eq6}
E=\frac{1}{2} \sum_{i, j, i \neq j} \phi_{i j}\left(r_{i j}\right)+\sum_{i} F_{i}\left(\rho_{i}\right)
\end{equation}
where $\phi_{i j}$ represents the pair energy between atoms $i$ and $j$ at separation $r_{ij}$, and $F_{i}$ is the embedding energy associated with embedding an atom $i$ into a position with an electron density $\rho_{i}$ and functional form $\phi(r)$ reads:

\begin{equation}
\phi(r)=\frac{A \exp \left[-\alpha\left(r / r_{e}-1\right)\right]}{1+\left(r / r_{e}-\kappa\right)^{20}}-\frac{B \exp \left[-\beta\left(r / r_{e}-1\right)\right]}{1+\left(r / r_{e}-\lambda\right)^{20}}
\end{equation}

where $r_{e}$ is the equilibrium spacing between nearest neighbours, $A, B, \alpha$, and $\beta$ are four adjustable parameters, and $\kappa$ and $\lambda$ are two additional parameters for the cutoff length. 
The electron density writes:
\begin{equation}
\rho_{i}=\sum_{j, j \neq i} f_{j}\left(r_{i j}\right)
\end{equation}
with $f_{j}\left(r_{i j}\right)$ the electron density at atom $i$ due to atom $j$ at distance $r_{i j}$ taking the form, with $f_e$ an adjustable parameter:
\begin{equation}
f_j(r_{ij})=\frac{f_{e} \exp \left[-\beta\left(r_{ij} / r_{e}-1\right)\right]}{1+\left(r_{ij} / r_{e}-\lambda\right)^{20}} .
\end{equation}
For a pure element $a$, the EAM potential is thus composed of three functions: the pair energy $\phi$, the electron density $\rho$, and the embedding energy $F$. For two interacting atoms $a$ and $b$ , the Johnson mixing rule is applied \cite{Johnson1989}, leading to pair potential:
\begin{equation}
\label{eq7}
\phi^{a b}(r)=\frac{1}{2}\left[\frac{f^{b}(r)}{f^{a}(r)} \phi^{a a}(r)+\frac{f^{a}(r)}{f^{b}(r)} \phi^{b b}(r)\right] .
\end{equation}

Embedding energy functions are defines by three equations. For a smooth variation of the embedding energy, these equations are required to match values and slopes at their junctions. 

\begin{align}
F(\rho) & =\sum_{i=0}^{3} F_{n i}\left(\frac{\rho}{\rho_{n}}-1\right)^{i}, \quad \rho<\rho_{n}, \quad \rho_{n}=0.85 \rho_{e}, \\
F(\rho) & =\sum_{i=0}^{3} F_{i}\left(\frac{\rho}{\rho_{e}}-1\right)^{i}, \quad \rho_{n} \leqslant \rho<\rho_{0}, \quad \rho_{0}=1.15 \rho_{e}, \\
F(\rho) & =F_{e}\left[1-\ln \left(\frac{\rho}{\rho_{s}}\right)^{\eta}\right]\left(\frac{\rho}{\rho_{s}}\right)^{\eta}, \quad \rho_{0} \leqslant \rho .
\end{align}

Extensions of EAM, valid for more systems, known as modified embedded atom method (MEAM) and 2nd nearest-Neighbor MEAM have been proposed \cite{Baskes1992, Lee2000} for improving accuracy.\\
These forcefields are widely used for plasma sputtering deposition, nanoparticle growth and plasma treatment of alloyed materials.\\

When plasma chemistry comes into play, like for plasma polymerisation or grafting \cite{Zarshenas2018, Brault2022, Jagodar2022, Kandjani2023}, plasma-liquid interactions \cite{Yusupov2014}, plasma-medicine studies \cite{Bogaerts2014, Neyts2014, Bogaerts2015}, nanoparticle growth in low-temperature plasmas \cite{Barcaro2019, Brault2019}, reactive (and variable charge) potentials are needed. Fortunately there are some available suitable forcefields, keeping in mind they are not always highly transferable. Preliminary tests are required for verifying applicability on basic properties of the system under study.

The most simple and faster reactive potential is the Reactive Bond Order Potential (REBO) that has been developed for carbon and hydrocarbon systems \cite{Brenner1989, Brenner1990, Brenner2002, Ni2004, Fonseca2011}.

The REBO (Reactive Empirical Bond Order) potentials is an extension of Tersoff potential \cite{Tersoff1988, Tersoff1988a, Tersoff1988b}. The modifications brought by Brenner concern improvements of bond order, repulsive and attractive pair terms. There are two generation of REBO Potential. In the first generation potential, the total energy of hydrocarbons writes:

\begin{equation}
\label{eq8}
E=\sum_{i} \sum_{j(>) i}\left[V_{\rm R}\left(r_{i j}\right)-\bar{B}_{i j} V_{\rm A}\left(r_{i j}\right)\right]
\end{equation}

Where the function $\bar{B}_{i j}, V_{\rm R}\left(r_{i j}\right)$ and $V_{\rm A}\left(r_{i j}\right) f_{c}\left(r_{i j}\right)$ are the bond order, repulsive and attractive potential terms, defined as:
\begin{equation}[
\bar{B}_{i j}=\left(b_{ij}+b_{j i}\right) / 2+F_{i j}\left(N_{i}^{(t)}, N_{j}^{(t)}, N_{ij}^{(conj)}\right)
\end{equation}

The interpolation function $F_{i j}$ is used to make the potential continuous, using the cutoff function $f_c(r)$:
\begin{align}
f_{c}= \begin{cases}1 & \text { if } r<R-D \\ \frac{1}{2}-\frac{1}{2} \sin \left[\frac{1}{2} \pi(r-R) / D\right] & \text { if } R-D<r<R+D \\ 0 & \text { if } r>R+D\end{cases}
\end{align}

 In this way the $N_{i}^{(t)}$ and $N_{j}^{(t)}$ which are the number of atoms, respectively, bonded to atom $\mathrm{i}$ and $\mathrm{j}$, defined the total number of neighbours. $N_{i j}^{(\text {conj })}$ is the conjugated term of atoms $\mathrm{i}$ and $\mathrm{j}$. 
\begin{equation}
N_{i}^{(t)}=\sum_{i(=t)} f_{c}\left(r_{i j}\right.
\end{equation}
Full details of each term are described by Brenner \cite{Brenner1990}.

Here the repulsive and attractive pair terms with new parameter are given by \cite{Brenner1990}:
\begin{align}
V_{R}\left(r_{i j}\right)& = f_{i j}\left(r_{i j}\right) D_{i j}^{(e)} /\left(S_{i j}-1\right) \text{exp} \left[-\sqrt{2 S_{i j}} \beta_{i j}\left(r-R_{i j}^{(e)}\right)\right] \\
V_{A}\left(r_{i j}\right)& = f_{i j}\left(r_{i j}\right) D_{i j}^{(e)} S_{i j} /\left(S_{i j}-1\right)\text{exp}\left[-\sqrt{2/S_{i j}} \beta_{i j}\left(r-R_{i j}^{(e)}\right)\right] 
\end{align}

The function $f_{i j}(r)$, which restricts the pair potential to nearest neighbours, is given by:
\begin{align}
f_{i j}(r)= \begin{cases}1 & \text { if } r<R_{i j}^{(1)} \\ {\left[1+\cos \left[\frac{\pi\left(r-R_{i j}^{(1)}\right)}{\left(R_{i j}^{(2)}-R_{i j}^{(1)}\right)}\right]\right] / 2,} & \text { if } R_{i j}^{(1)}<r<R_{i j}^{(2)} \\ 0 & \text { if } r>R_{i j}^{(2)}\end{cases}
\end{align}
This form makes the correspondence to Morse functions more apparent. If $S_{i j}=2$, then the pair terms reduce to the usual Morse potential. Furthermore the depth parameter $D_{i j}^{(e)}$, equilibrium distance $R_{i j}^{(e)}$ and $\beta_{i j}$ are equal to the usual Morse parameters.

Since first generation is not considering the different types of bonding like triple, double or single bonds, the second generation REBO has been developed \cite{Brenner2002}, by introducing a generalization of the bond order function $B_{i j}$, written as:
\begin{equation}
\label{eq9}
\bar{B}_{i j}=\frac{1}{2}\left[b_{i j}^{\sigma-\pi}+b_{j i}^{\sigma-\pi}\right]+b_{j i}^{\pi}
\end{equation}
The functions $b_{i j}^{\sigma-\pi}$ and $b_{j i}^{\sigma-\pi}$ depend on the local coordination and the bond angle for atoms $\mathrm{i}$ and $\mathrm{j}$.The function $b_{ji}^{\pi}$ is further written as:

\begin{equation}
b_{ji}^{\pi}=\prod_{i j}^{RC}+\prod_{i j}^{DH}\\
\end{equation}

Where $\prod_{i j}^{RC}$ depend on the bond whether the conjugate bond $i$ and $j$ is a radical character. And the term $\prod_{i j}^{D H}$ depends on the dihedral angle for carbon-carbon double bonds. The  term $b_{i j}^{\sigma-\pi}$ in Equation~(\ref{eq9}) is given by:

\begin{equation}
b_{i j}^{\sigma-\pi}=\left[\sum_{k \neq i, j} f_{i k}^{c}\left(r_{i k}\right) G\left(\cos \left(\theta_{i j k}\right)\right) \exp \lambda_{i j k}+P_{i j}\left(N_{i}^{(C)}, N_{j}^{(H)}\right)\right]^{-\frac{1}{2}}
\end{equation}

where $f_{i k}^{c}$ is the cutoff function ensures that the interactions include nearest neighbours only. the function $\mathrm{P}$ represents a bicubic spline for interpolation of the potential.  $G\left(\cos \left(\theta_{i j k}\right)\right)$ is the angular function. The quantities $N_{i}^{(C)}$ and $N_{i}^{(H)}$ represent the number of carbon and hydrogen atoms for hydrocarbon species, respectively, that are neighbours of atom $i$. They are defined by sum:

\begin{align}
N_{i}^{(C)} & =\sum_{k \neq i, j}^{\text {carbon,atoms }} f_{i k}^{c}\left(r_{i k}\right) \\
N_{i}^{(H)} & =\sum_{l \neq i, j}^{\text {hydrogen,atoms }} f_{i l}^{c}\left(r_{i l}\right)
\end{align}

For adding more flexibility and accounting of long range interactions between hydrocarbon species, the 2nd generation REBO potential has been extended by adding a torsional and Lennard-Jones potential, leading to the AIREBO (Adaptive Intermolecular (Reactive Empirical Bond Order) potential. In this case the total AIREBO energy of system is:
\begin{equation}
E=E^{REBO}+E^{LJ}+E^{\text {tors }}
\end{equation}

Where the term $E^{LJ}$ is the Lennard-Jones potential, contributing to the energy. It ensures the interaction at large distance making AIREBO an intra and inter molecular potential.

$E^{\text {tors }}$ is the torsion contribution, needed for studying reactivity of large hydrocarbon molecules.

For going beyond hydrocarbon molecules, two other forcefields have been developed and are among the most used: ReaxFF \cite{Duin2001, Chenoweth2008, Senftle2016, Liang2013a} and COMB (Charge Optimized Many Body) \cite{Shan2010,Liang2013b, Liang2013a}.
The bond order is defined as distance dependent, for precisely describing bond formation and breaking. They add the possibility of varying the charge of each atom \cite{Rappe1991, Rick1994}. Variable charge concept is based on electronegativity equalization following three assumptions: (a) the electronegativity of an atomic site is dependent on the atom’s type and charge and is perturbed by the electrostatic potential it is subjected from neighbours (b) charge transfers between atomic sites respect electronegativity equality. (c) The variable charges obey an extended Lagrangian equation in which they have a fictitious mass, velocities, and kinetic energy and then moved with respect to Newtonian mechanics.

In the ReaxFF potential, the total energy of a system is given by a summation of all contribution of interaction on the system; this ReaxFF overall system energy is given by \cite{Duin2001}:

\begin{align}
E_{\text {system }} = & E_{\text {bond }}+E_{L p}+E_{\text {over }}+E_{\text {under }}+ \nonumber \\
&E_{\text {val }}+E_{\text {pen }}+E_{\text {coa }}+E_{c_{2}}+E_{\text {tors }}+  \\
&E_{\text {conj }}+E_{H-\text { bond }}+E_{v d \text { Waals }}+E_{\text {Coulomb} \nonumber}
\end{align}

Where:

$E_{\text {bond}}$ is the bond order energy $E_{L p}$ is the Lone pair energy

$E_{\text {over}}$ is overcoordination energy

$E_{\text {unde}}$ is undercoordination energy

$E_{val}$ is valence angle term

$E_{pen}$ is penalty energy

$E_{coa}$ is three-body conjugation term

$E_{c_{2}}$ is Correction for $C_{2}$

$E_{tors}$ is torsion angle terms

$E_{\text {conj }}$ is four body conjugation term

$E_{H-bond}$ is Hydrogen bond interactions

$E_{vdWaals}$ is van der Waals interactions

$E_{\text {coulomb}}$ is Coulomb interaction.

At this step, the focus is on bond order and for maintaining clarity, the details of the various energy terms are not provided here, but can be found elsewhere \cite{Duin2001, Chenoweth2008} with a great detail.

The term of bond order energy is the most developed in reaxFF and is a sum of three terms: single bond, double bond and the triple bond. $BO_{ij}$ is defined as:

\begin{equation}
BO_{ij}=BO_{ij}^{\sigma}+BO_{ij}^{\pi}+BO_{ij}^{\pi\pi}
\end{equation}
Where
\begin{align}\label{eq10}
BO_{ij}^{\sigma} & =\exp \left[b_{bo,1}\left(\frac{r_{ij}}{r_{0}^{\sigma}}\right)^{p_{bo,2}}\right] \\
BO_{ij}^{\pi} & =\exp \left[p_{bo,3}\left(\frac{r_{ij}}{r_{0}^{\pi}}\right)^{p_{bo,4}}\right] \\
BO_{ij}^{\pi \pi} & =\exp \left[p_{bo,5}\left(\frac{r_{ij}}{r_{0}^{\pi\pi}}\right)^{p_{bo,6}}\right]   
\end{align}

where $B O_{i j}$ is the bond order between atoms $i$ and $j$, it depends on the local environment. For carbon-carbon interactions, all contributions in set of Equation~(\ref{eq10}) are used, leading to a max bond order of 3, while for C-H interaction only the $\sigma$ contribution is used, leading to a maximum bond order of 1 \cite{Chenoweth2008}.

For COMB family, there are 2 generations COMB \cite{Shan2010} and COMB3 \cite{Liang2013b}. They mainly differ by the atom type involved. It is a similar approach to reaxFF potential, providing with a complementary atom selection. For both reaxFF,COMB and COMB3, there is no predefined molecule but atom assembly connected by interactions that forms (or not) molecule(s). A comparison between these forcefields is available\cite{Liang2013a}.

There are many other forcefields that can be used, either 2-body (pair potentials), 3-body and many-body, reactive or not. It is impossible to include a full list here. The main important issue is the force field parametrization that is often performed using various, experimental or theoretical (DFT calculated for example)  materials, molecule parameters. The larger the parametrization database is, the larger is the range of validity of the force field. Force field parameters are available as files from supplementary information or database like NIST Interatomic Potentials Repository (https://www.ctcms.nist.gov/potentials/) or OpenKIM (https://openkim.org/). When unavailable, they should be determined from fitting procedures on experimental and calculated relevant quantities. These procedures might be very time consuming especially for reactive variable charge force fields. 

\subsection{Selected MS simulations tools}
Handling MD simulations for plasma applications require softwares able to run using parallel computation coding and facilities. This is necessary since either there is a large number of species, for reaching statistical meaning or because force field are enough complex (reaxFF, COMB3) for requiring enough CPU ressources.
The most popular multipurpose sofware is LAMMPS \cite{LAMMPS} (https://lammps.org). It is able to address numerous problem in materials science as well as liquid and gaseous. A huge number of force field parameter files are readable by LAMMPS. A large number is already available in the potentials folder of the distribution. It also includes recent machine learning potentials. Regular version updates allow to solve bug issues as well as adding new potentials and fonctionnalities. Statistical quantities can be computed such as X-ray diffraction patterns of films or plasma treated bulk systems, as well as thermal conductivity. Pre-and post-processing tools are also listed on the website. Some of them being directly able to provide input data in LAMMPS format as well as reading ouput LAMMPS data for calculating statistical quantities. Main interest is LAMMPS is fully free and open-source. It is running on operating systems: Windows, Linux and Mac. Many tutorial are availble too.
A friendly and interactive user forum is available at https://matsci.org/c/lammps/. Most of the new user questions are already answered.
Similar to LAMMPS, DL$\_$POLY \cite{Todorov2006} is also available for download at http://www.ccp5.ac.uk/DL$\_$POLY/. It is free for academic users, as well as NAMD \cite{Phillips2020}, rather designed for MD in the field of structural biology, but with possible relevance for plasma-mdecine applications.
There also exist numerous commercial softwares suitable for plasma applications. Among them Materials Studio (https://www.3ds.com/products-services/biovia/products/molecular-modeling-simulation/biovia-materials-studio/) and AMS suite from SCM company (http://www.scm.com) are offering advantageous possibilities for addressing plasma processing.

\section{MD simulations for plasma processing}\label{sec3}
It is clear that MD simulations are able to treat interaction between neutral atoms and molecules. It only requires the best forcefield and relevant initial conditions. In plasma, ions are often treated as fast neutral, especially for plasma deposition processes. Recently, accounting the ion potential energy has been achieved adding a repulsive short range potential between HiPIMS generated depositing ions and surface atoms \cite{Kateb2021}. But the charge can be explicitly given and a charge dependent potential is included. Either the long range is treated in direct space with a large cutoff distance or long range part is treated separately in the k-space. This also requires good forcefields. Reactive variable charge potentials described in Section~\ref{subsec24} are also usable when necessary, but at expense of higher computer time than other constant charge forcefields.
This section will focus on new progress either for processes now tractable by MD simulations or emerging/hot/complex plasma topics 

\subsection{MD simulations of ``new'' plasma processes}\label{subsec31}
A way for including plasma effect is the addition of an electric field in the MD simulations. As examples, electric field effects has been studied for monitoring carbon nanotube (CNT) growth \cite{Neyts2011}, pore formation in plasma interaction with phospholipid layers (as encountered in plasma medicine studies) \cite{Yusupov2017}.

\begin{figure}[h]%
\centering
\includegraphics[width=0.9\textwidth]{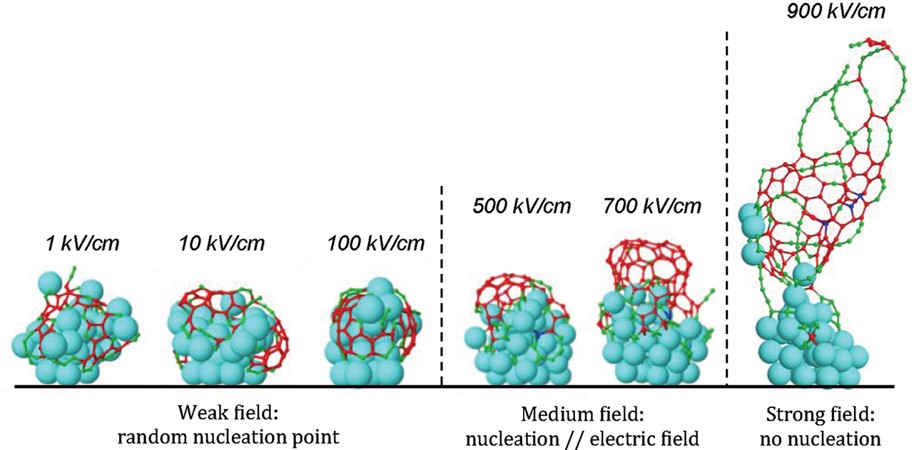}
\caption{Effect of applying an electric field on the nucleation of a SWNT cap. The small red atoms are 3-coordinated carbon atoms, the small green atoms are 2- or 1-coordinated carbon atoms. The large blue atoms represent nickel atoms. Reprinted  from Neyts et al. \cite{Neyts2011} with permission. Copyright {2011} American Chemical Society. }\label{fig2}
\end{figure}

In the first case, a supported Ni catalyst is exposed to carbon vapour a a given temperature and a constant electric field is applied throughout the deposition process. The CNT growth  is thus monitored by this electric field.  It is observed that three electric field regimes are effective: weak (1-100 kV.cm$^{-1}$), medium (500-700 kV.cm$^{-1}$) and strong (900 kV.cm$^{-1}$) fields.  In the weak field regime the growth is operating through random nucleation as without electric field. In the medium field regime, nucleation is parallel  to the electric field vector while for strong field no nucleation on the catalyst occurs, only random bond between carbon atoms. Figure~\ref{fig2} summarizes the CNT growth vs electric field magnitude.\\

\begin{figure}[h]%
\centering
\includegraphics[width=0.9\textwidth]{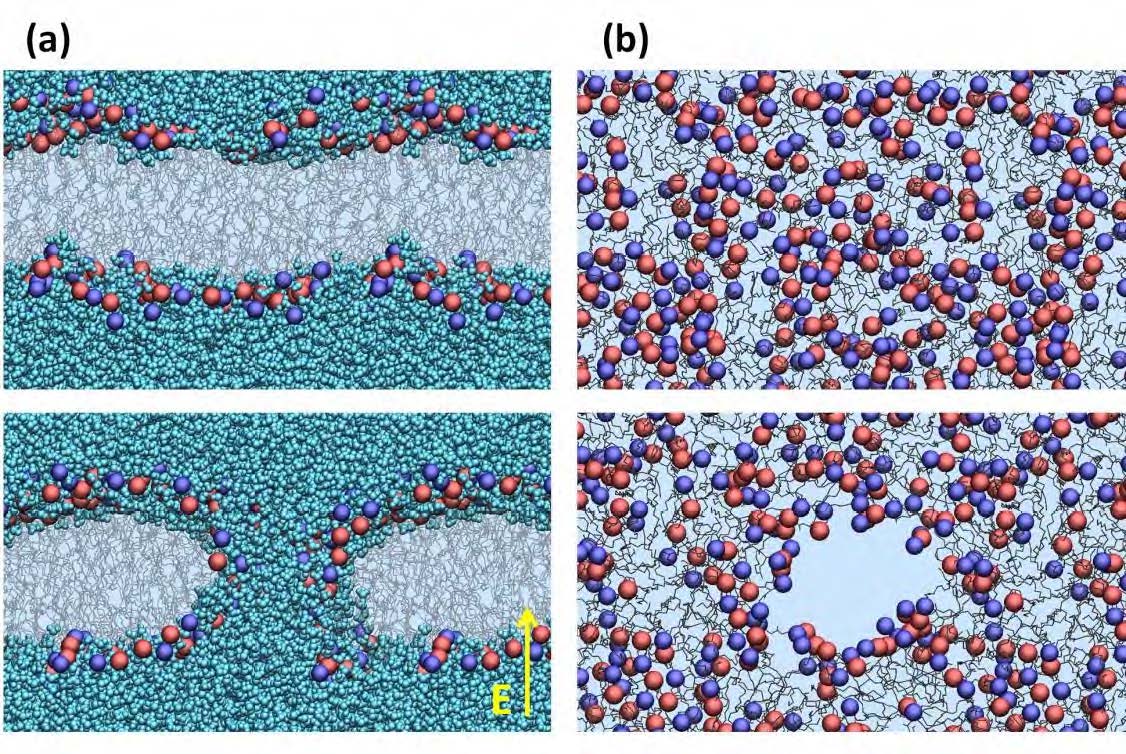}
\caption{Snapshots from MD simulations, showing the pore formation in a native PLB after $\approx$ 2 ns, upon effect of a constant electric field of 0.5 V.nm$^{-1}$, (a) side view and (b) top view. The water layers are removed from the top view picture, for the sake of clarity. Reprintedfrom Yusupov et al. \cite{Yusupov2017} with permission . Copyright {2017} Elsevier. }\label{fig3}
\end{figure}

When looking at plasma-medicine/biology applications, introducing electric field in addition to the reactive species (the so-called RONS, Reactive Oxygen and Nitrogen Species \cite{Morabit2021}) is of paramount importance as demonstrated, for example, by the plasma interaction with phospholipid bilayers (PLB). In this case applying an electric field (0.5 V/nm) results in the formation of pores in  the BLP as shown in Figure~\ref{fig3}. These pores are expected to facilitate the delivering of ROS (Reactive Oxygen Species) to the PLB, and thus accelerating oxidation and possible damages \cite{Yusupov2017}. Increasing the electric field reduces the formation time of the pores, and thus accelerates the reactivity of ROS with BLP.

\begin{figure}[h]%
\centering
\includegraphics[width=0.9\textwidth]{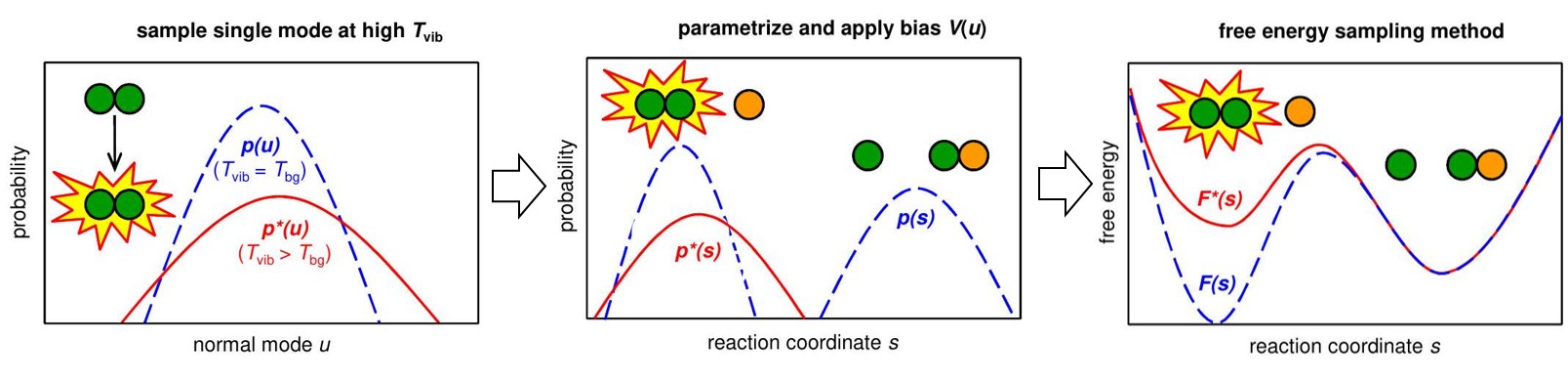}
\caption{Overview of the principles behind the approach to vibrational excitation. At high $T_{\text {vib }}$, the probability distribution $p^{*}(\boldsymbol{u})$ along the normal mode differs from the equilibrium distribution $p(\boldsymbol{u})$ at $T_{\mathrm{bg}}$. As a result, the probability distribution along a reaction coordinate $\boldsymbol{s}$ is also affected, which leads to a change in the apparent reaction free energy barrier. In our method, this modified $F^{*}(\boldsymbol{s})$ (or $\left.p^{*}(\boldsymbol{s})\right)$ is obtained from a free energy simulation after applying a bias potential $V(\boldsymbol{u})$ that enforces $p^{*}(\boldsymbol{u})$ at $T_{\mathrm{bg}}$. Adapted  from Bal et al.  \cite{Bal2019} with permission. Copyright {2019} American Chemical Society.}\label{fig4}
\end{figure}

Another mechanism, not taken into account until now, has been successfully addressed recently: It is including vibrational excitation in MD simulations. Vibrational excitation is a quantum concept. Since there is no vibrational quantum number in classical mechanics, it is not possible to populate and keep vibrational levels in the course of the MD simulations. To circumvent this impossibility, a very interesting idea \cite{Bal2019} was to apply a bias potential to the vibrational energy accordingly to the probability distribution at the excited  temperature $T_{vib}$. So the strategy is to model systems in which most modes are in equilibrium with each other at a background temperature ($T_{bg}$, say 300K for example), while certain selected modes have a (higher) vibrationally excited temperature ($T_{vib}$). So, probability distribution $p(\vec{R})$ of any system in configuration space at temperature $T$ and potential energy $U(R)$ follows the Boltzmann distribution, used for each two temperatures: $ p(\vec{R}) \propto e^{-\frac{U(\vec{R})}{k_BT}}$ which leads to the potential energy surface along the reaction coordinate $s$, $F(s) = -k_BT\ln p(s) + C$. The change in the potential energy curves is illustrated in Fig.~\ref{fig4}. 

A remaining major question arising in MD simulations of low temperature plasma is the explicit inclusion of electron motions in Newton equations. If it is widely done in warm dense (plasma)  matter through the use of eFF forcefields, which a special case of Wave-packet molecular dynamics \cite{Su2007, J-Botero2011, Lavrinenko2019,Davis2020, Jin2021} or screened potential describing charged particles with ions and atoms \cite{Ramazanov2002, Ramazanov2010}. The question for extending the use of this potential, and in which way, in low-pressure low-temperature plasmas, remains open.

\begin{figure}[h]%
\centering
\includegraphics[width=0.9\textwidth]{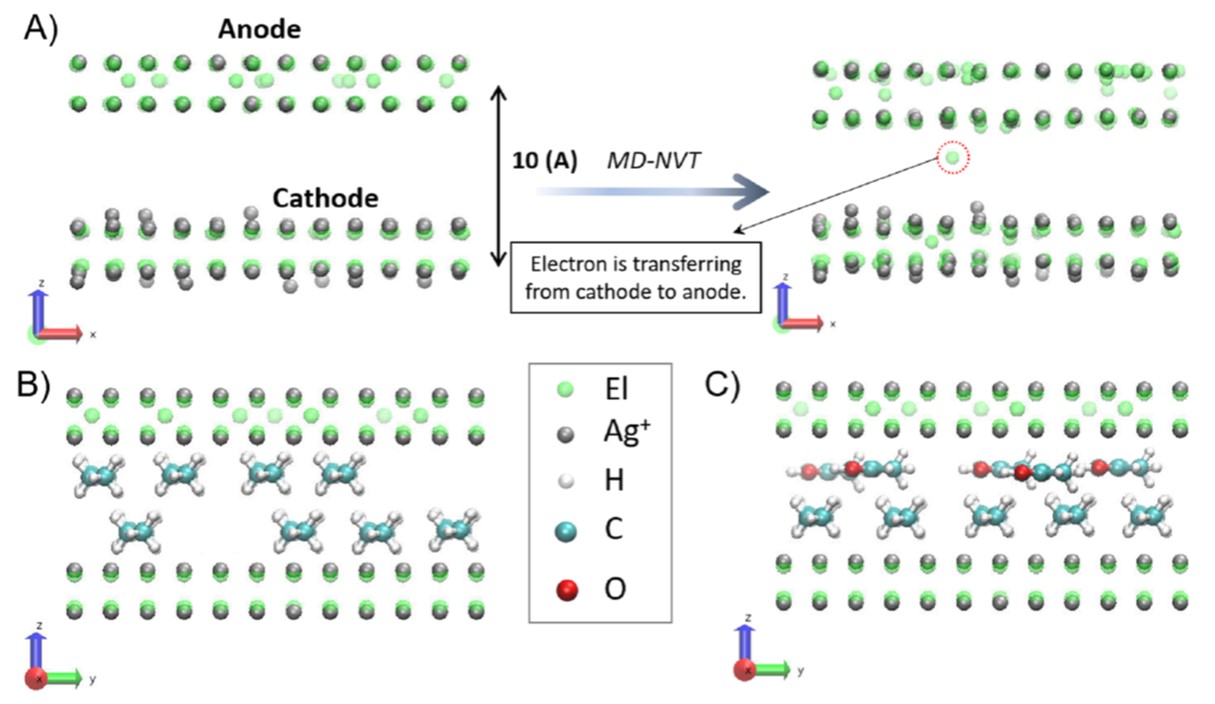}
\caption{(a) Two 6 × 6 silver slabs including 72 Ag$^+$ and 72 electrons. The slabs were separated by 10 \AA, and 10 electrons were transferred from the bottom layer to the top layer to apply 40.7 V electric potential to the system. The time that the first electron started transferring from the anode to the cathode is defined as the TDDB. (b) Two silver slabs and eight decane molecules added into the vacuum space between the cathode and the anode. (c) Two silver slabs with five acetophenone and five decane molecules added into the vacuum space between the cathode and the anode. Reprinted  from Akbarian et al. \cite{Akbarian2021} with permission. Copyright {2021} American Institute of Physics.}\label{fig5}
\end{figure}

\begin{figure}[h]%
\centering
\includegraphics[width=0.9\textwidth]{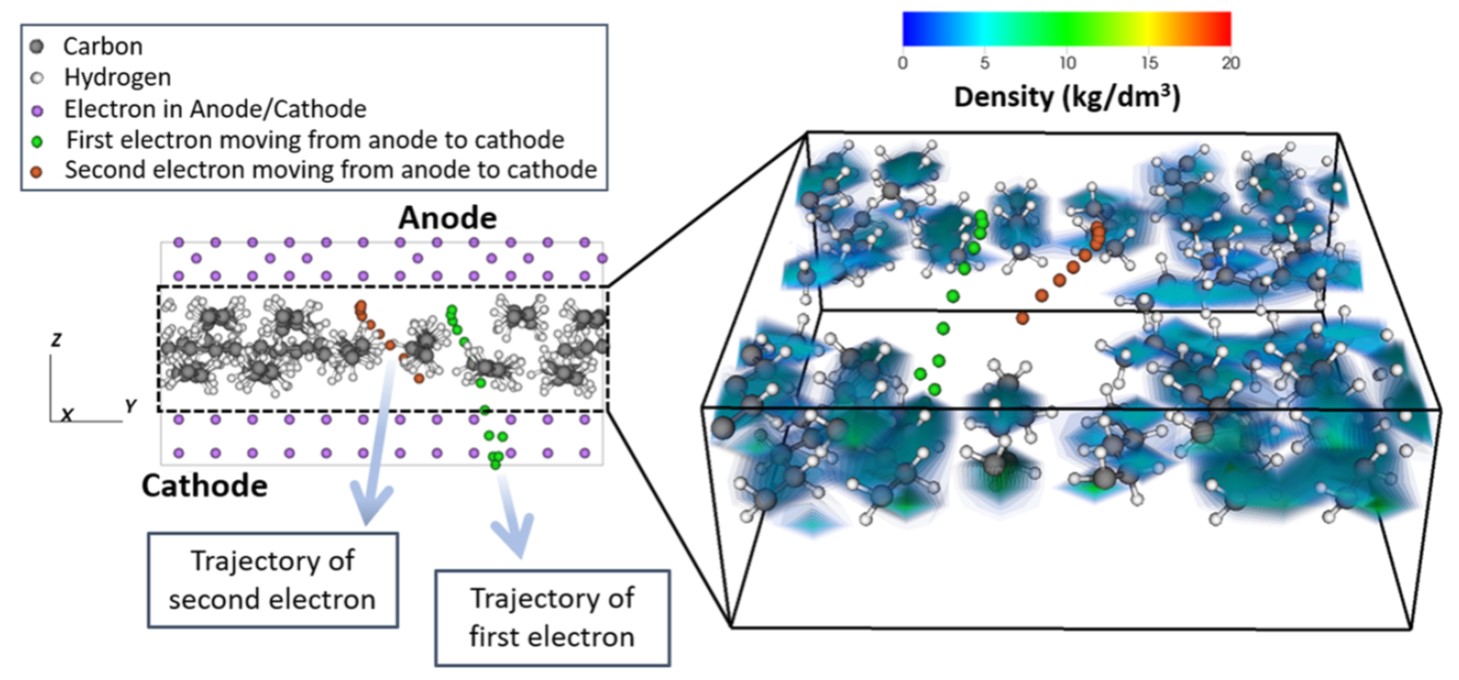}
\caption{Contour of the system density between the cathode and the anode and the trajectory of the first two electrons transferring from the anode to cathode, indicating that electrons traverse through the voids during the electrical breakdown.. Reprinted  from Akbarian et al. \cite{Akbarian2021} with permission. Copyright {2021} American Institute of Physics.}\label{fig6}
\end{figure}

Very recently, the force field e-reaxFF was parametrized for allowing description of discharge breakdown between two Silver electrodes separated by an insulating polymer \cite{Akbarian2021} (Figure~\ref{fig5}). It allows to follow the electron trajectories. One electrode is supporting electrons and when running the MD simulation, the electrons migrate towards the counter-electrode as shown in Figure~\ref{fig6}, for which trajectories are evolving along void channels between polymer molecules. 

Moreover the electric breakdown is shown to occur after a decreasing delay time when increasing applied voltage magnitude. But, effects of considering $m_e$ = 1 should be further investigated. It should also kept in mind that using e-reaxFF, with other materials than Silver for studying electrical breakdown, requires a new force field parametrization.

This pioneering and breakthrough work open the way to explicitly include electrons in MD simulations in the context of low temperature plasma physics and chemistry, especially for producing high-throughput electron collision data in any situations.

\subsection{MD simulations for plasma applications}\label{subsec32}

\begin{figure}[h]%
\centering
\includegraphics[width=0.9\textwidth]{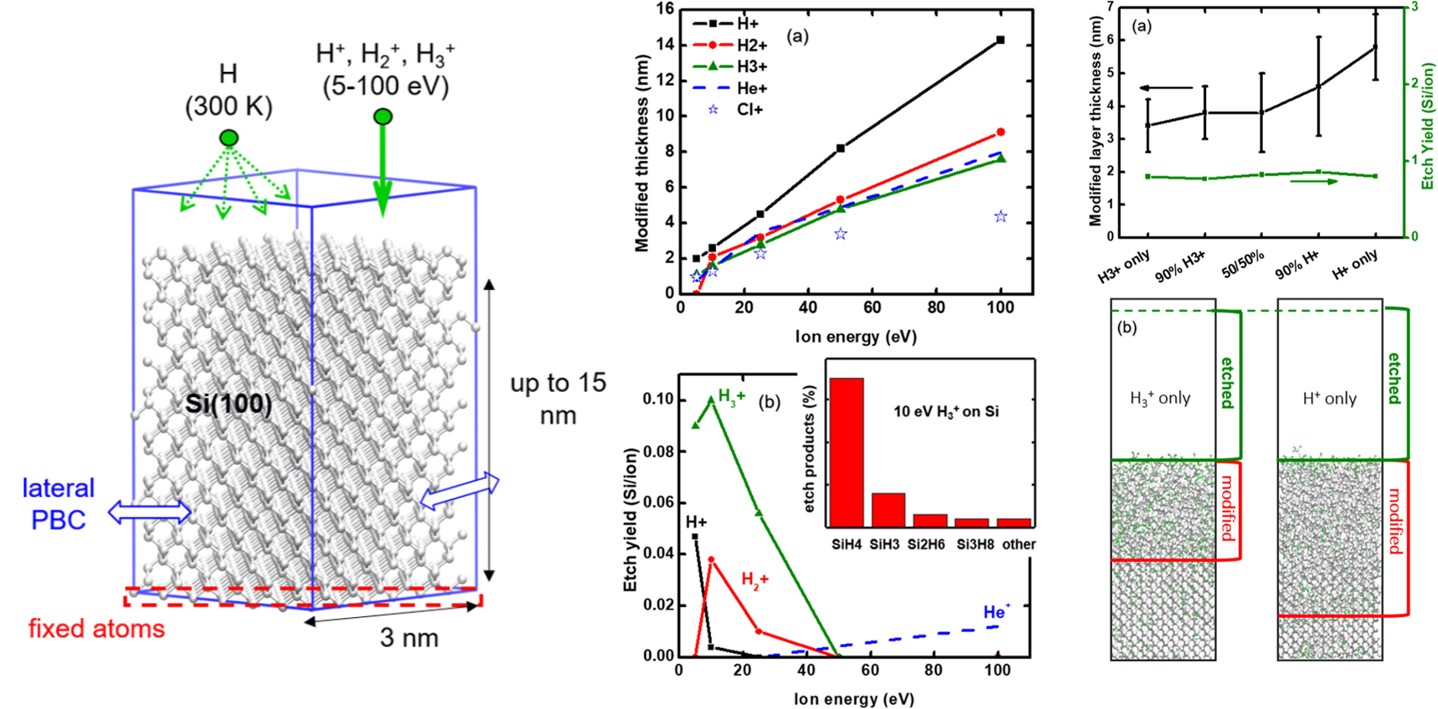}
\caption{(Left part) Initial (100) c-Si simulation cell used in the MD calculations, (central part) H$^+_x$ ion bombardment of silicon. (a) Thickness of the modified layer at steady state as a function of the ion energy, for different ion types. Values for pure He$^+$ \cite{Martirosyan2018} and Cl$^+$ \cite{Brichon2014} ion bombardment were added for comparison. (b) Etching yields (EY)  at steady state as a function of the ion energy, for different ion types. The distribution of etch products for a 10 eV H$_3^+$ bombardment is shown in the inset graph. (right panel) Mixed H$^+$ ion/H radical bombardment of Si (ion energy $E_{ion}$ = 100 eV, radical to ion flux $\Gamma$ = 10) (a) Modified layer thickness and EY, at steady state, as a function of the ion composition. (b) Snapshots of the cells corresponding to the $H^+$ only and $H^+$ only cases for an ion dose $\approx$ 3.5 x 10$^{16}$ ion cm$^{-2}$.  Adapted  from Martisoryan et al. \cite{Martirosyan2019} with permission. Copyright {2019} Institute of Physics.}\label{fig7}
\end{figure}

Microelectronics is an historical playground for low-temperature plasma physics and chemistry. It has, without any doubt, driven many progresses in the field. Very recently atomic scale processes have been gain a huge interest due to the miniaturization effort driven by increasing computing performances on smaller and smaller devices. Very recently, an account of atomistic simulation, besides experiments, has been reviewed for almost all processes of microelectronics: etching processes, atomic layer deposition, etc \cite{Ishikawa2019}. The level of reachable details is now very impressive. For example, due to available robust and performing interaction potentials, MD simulations of etching processes can be closely connected to experiments. In a recent work \cite{Martirosyan2019}, etching with H, H$^+$,   H$_2^+$ and H$_3^+$ species, is analysed. Especially, when considering Hydrogen ions with different masses, same etch rates are obtained while Silicon affected zone is thicker with low mass ions. Figure~\ref{eq7} summarizes the main features of Hydrogen etching of silicon.\\

\begin{figure}[h]%
\centering
\includegraphics[width=0.9\textwidth]{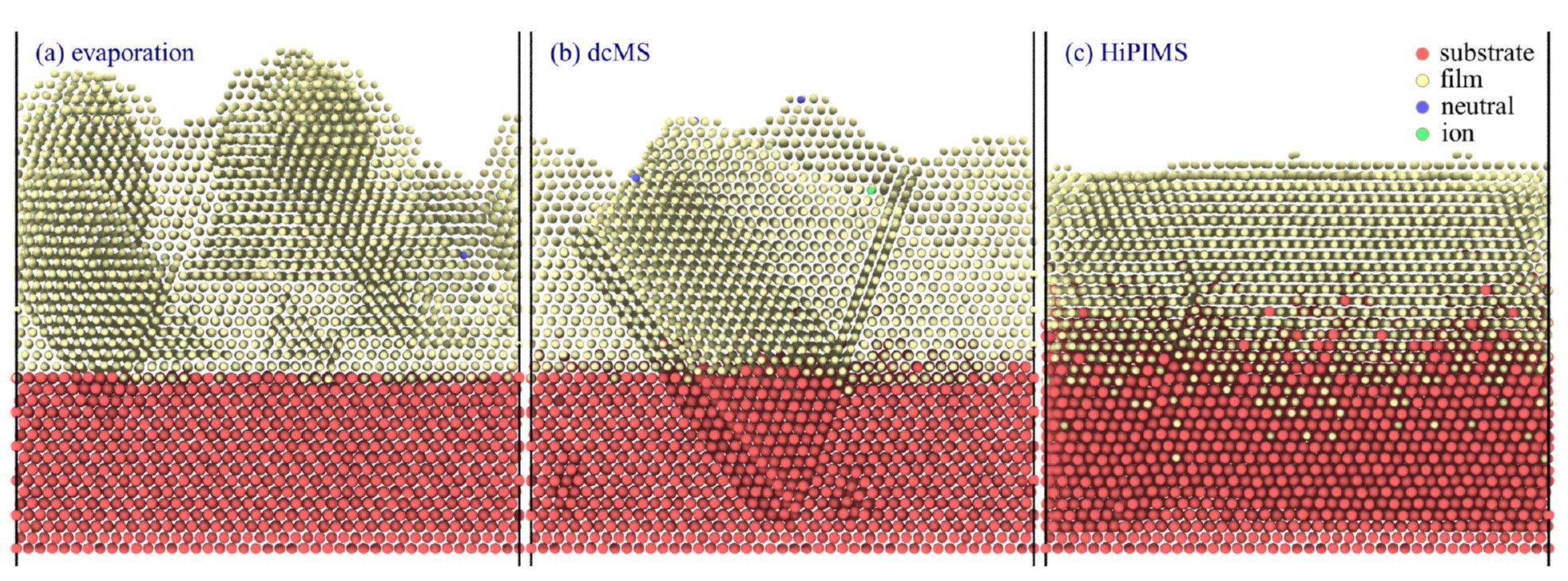}
\caption{(Illustration of interface mixing using (a) thermal evaporation, (b) dcMS and (c) HiPIMS after 2.5 ns deposition. The red, green, blue and yellow are indicating substrate, neutral, ions and lm atoms.. Reprinted from Kateb et al. \cite{Kateb2019} with permission. Copyright {2019} American Vacuum Society}\label{fig8}
\end{figure}

\begin{figure}[h]%
\centering
\includegraphics[width=0.9\textwidth]{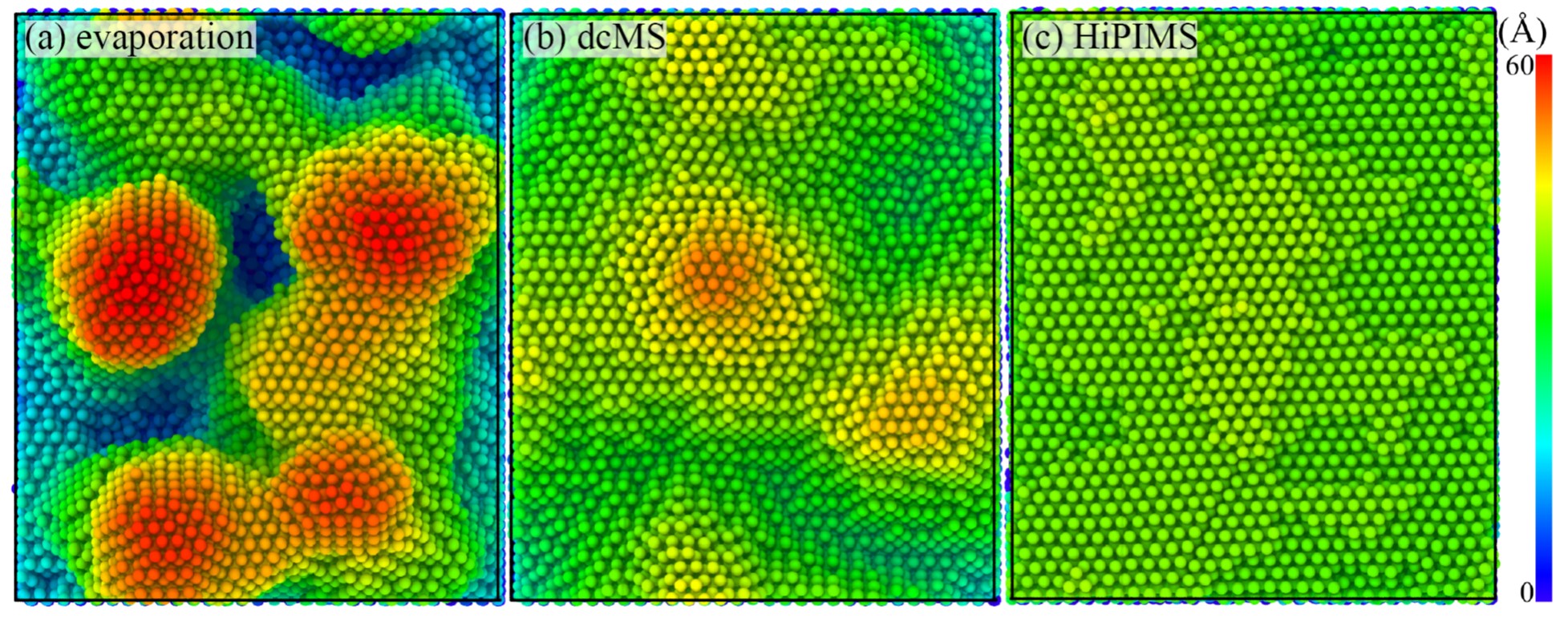}
\caption{The surface topology obtained using (a) thermal evaporation (b) dcMS and (b) HiPIMS deposition with similar deposition time and energy distribution. The deep blue indicates substrate surface and red denotes thickness higher than 6 nm.. Reprinted from Kateb et al. \cite{Kateb2019} with permission. Copyright {2019} American Vacuum Society}\label{fig9}
\end{figure}

A plasma process which meets more and more interest is High Power Impulse Magnetron Sputtering (HiPIMS) deposition, especially for designing complex alloy coatings for many applications \cite{Anders2017,Gudmundsson2022}. A main feature of HiPIMS is to produce impinging fast metal ions on the surface substrate to be coated. A close comparison of the sputtered ion effects on coatings properties from different plasma sputtering processes and thermal evaporation has been recently reviewed \cite{Kateb2019,Kateb2021}. The main difficulty is to account for the high energy part of sputtered atom energy distribution function. Accounting it as a potential energy allowed to reproduce the main features of deposited films. Moreover, predictions for thermal evaporation and conventional dc sputtering are also well recovering experimental findings. Figures~\ref{fig8} and \ref{fig9} display the comparison of resulting  microstructure of the simulated film deposition for the three processes.

Clearly, HiPIMS process provides the best roughness and film-substrate interface compatible with good adhesion. This also depends on the sputtered ion to neutral ratio.

Further improvement of the method might benefit from using experimental energy resolved high resolution mass spectrometry of both neutral and ions. \\

\begin{figure}[h]%
\centering
\includegraphics[width=0.9\textwidth]{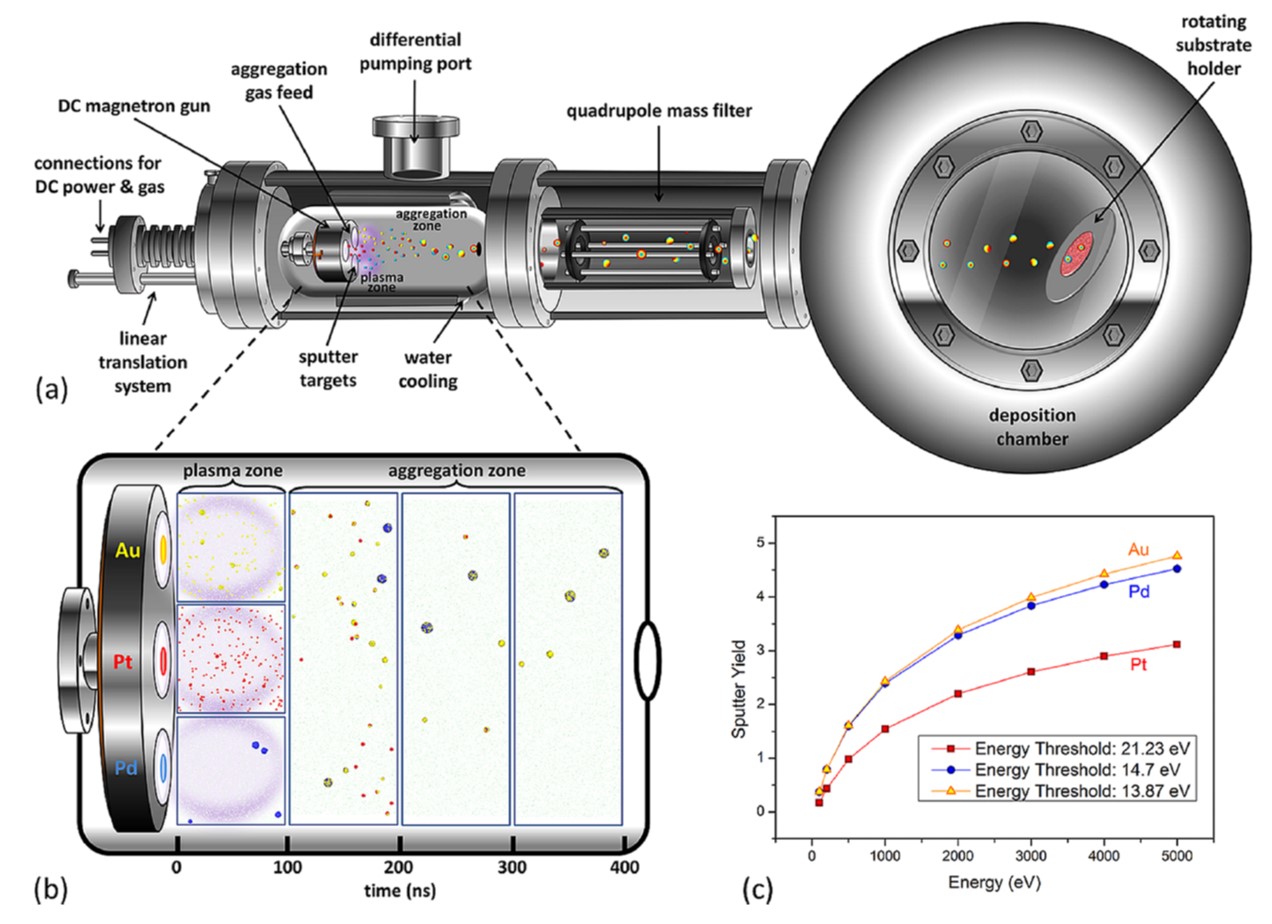}
\caption{T(a) Schematic representation of a magnetron-sputtering inert-gas-condensation system utilizing a triple-target configuration. (b) Schematic representation of the MD arrangement and its correspondence to the experimental setup. For the first 100 ns, individual nucleation of each element within plasma zones was simulated in a 1000 K Ar gas environment inside 50 × 50 × 50 nm$^3$ simulation boxes. Next, growth within a room temperature aggregation zone represented by a single 150 × 50 × 50 nm3 simulation box was simulated for 300 ns. Au, Pt, Pd, and Ar atoms are represented by yellow, red, blue, and green spheres, respectively. (c) Calculated sputter yields for all three single-element targets for various energies, used as input for MD: Au and Pd have consistently similar yields, whereas the Pt yield is significantly lower, due to its higher sputter energy threshold.. Reprinted  from Grammatikopoulos et al \cite{Grammatikopoulos2016} with permission. Open Access data}\label{fig10}
\end{figure}

Nanoparticle growth is a vast field of applications that magnetron sputtering gas aggregation technique (GAS) is contributing \cite{Xirouchaki2004,Quesnel2010,Caillard2015, Kylian2023}. Basically, sputtering of a target is done in a vapour at enough high pressure ($> 10$ Pa) so that collisions of sputtered atoms in the vapour allow the gas phase condensation of sputtered atoms as nanoparticles. 
This atomic process is well suited for MD simulations. Primary attempts, since the pioneering work of Haberland \cite {Haberland1995} concerned Fe clustering in an Argon gas, thus mimicking gas aggregation \cite{Lummen2004}.  This was followed by modelling alloy nanoparticle growth in GAS for direct comparison with experiments \cite{Grammatikopoulos2016, Brault2019a,Mattei2019}. A relevant procedure is summarized in Figure~\ref{fig10}.\\

Gas phase plasma chemistry is also a nice field for investigating ion - neutral and neutral - neutral reactions. Making use of Equations~(\ref{eq2})-(\ref{eq4}) allows to well define the simulation box for direct comparison with experiments. The main difficulty is the knowledge of initial composition of the vapour. This can be achieved by experimental mass spectrometry or numerical kinetic/fluid models, as for hydrocarbon plasmas \cite{Kandjani2023}. The dependence of the MD simulation results, i.e. simulated mass spectra, polymerisation mechanisms, on these initial conditions, remains an open question. It certainly requires a large parametric study for classifying MD simulation results along experimental parameters. This will be necessary for assessing a correct comparison with experimental results. The same problem still arises for polymer growth at surfaces: an initial composition does not lead necessarily to a film that corresponds to experiments. The correspondence can unfortunately be unique. When comparing film characterisation, infrared (IR) spectra for example, simulated spectra can reproduce IR peak positions but it is more difficult to obtain peak ratios in agreement with experiments. Here also a parametric MD study is necessary for determining which parameters are affecting the simulated deposited films\cite{Brault2022}.\\

Molecular dynamics simulations in the context of Plasma-Liquid interactions \cite{Vanraes2018} is becoming a hot topic where two fields are very active: plasma-medicine/biology \cite{Bogaerts2014,Yusupov2015,Yang2023} and plasma treatment of wastewater \cite{Brault2021,Ghasemi2022,Zeng2022}. 
In both cases a first approach is to analyse, via MD simulations, the interactions of ROS, mainly HO$^{\bullet}$, O$^{\bullet}$ with biological materials and organic pollutants molecules in order to predict oxidation reaction pathways and formed products. The key point for such studies is the availability of reactive forcefields. Fortunately the above-mentioned reaxFF is widely used for such studies. Recently, a reaxFF parametrization of organochlorine molecules \cite{Wolf2022} will broaden the range of  emerging pollutant molecules that can be studied using reactive MD simulations.
Despite the great success of using reaxFF forcefields, it should noticed, that sometimes reaxFF exhibit energy barrier along reaction coordinates, larger than quantum chemistry predictions \cite{Zarshenas2018, Wolf2022}. A possible workaround is to increase the operation temperature up to a few additional hundred Kelvin. In this case, care should be taken to the offset value to correctly correlate with experiments, especially when calculating reaction rates. \\
Since, in both cases, plasma-medicine and wastewater treatment, water is always present, it is advantageous to carry out simulations where interacting species are surrounded by water molecules, preferably at usual water density (1g.cm$^{-3}$). It can slightly increase computer time, but secondary reactions can be allowed, thus increasing better description of reactions in the real environment.\\
There are at least two approaches, that can be relevant, if direct comparison with experiments is targeted (instead of conducting parametric studies). First of all, creating a simulation box containing a solution with water (or other liquid), the molecule of interest and a number RONS (in agreement with the expected or experimentally available ratio of RONS to this molecule). Then run MD simulation with a temperature ramp for identifying the possible oxidation and degradation products. Thus determining a temperature range realizing a reaction might help for designing the corresponding experimental process\cite{Brault2021}.\\
Another way consists in filling the simulation box with water and the molecule of interest and then periodically randomly injecting oxidative radicals (for mimicking RONS delivery to liquid in non-thermal plasma experiments), and identifying degradation products as well as calculating reaction products\cite{Brault2023}.

Despite different conditions compared to usual low temperature plasma, it is interesting to mention thermonuclear fusion which make use of MD simulations that could be useful for other plasma communities. It especially address in particular to study the plasma-facing materials in contact with the "cold" plasma sheath (scrape-off-layer). Some recent simulations include plasma specificities (high flux, energetic ions, high temperature...) for sputtering calculations \cite{Hodille2020}, dust growth \cite{Matuska2019,Niu2022}, atom implantation and diffusion \cite{Pentecoste2016,Pentecoste2017,Xiang2022,Setyawan2023} or reactivity of oxide layers \cite{Jelea2006}.

\section{Conclusion}\label{sec4}
During the past decade, the use of Molecular Dynamics (classical or ab-initio) has grown for addressing numerous fields of plasma physics and chemistry. The advent of high performance computers, accurate forcefields and  easy to use software (the list is too long and is not detailed here), both free and commercial, allow to address complex phenomena as those encountered in plasma volume and/or in interaction with materials and liquids. Almost all processes of plasma can now be described using MD simulations. A recent achievement is including vibrational excitation which opens the way to elucidate numerous mechanisms in the field of plasma catalysis, and more generally for any applications involving plasma chemistry. Recent explicit inclusion of electrons for describing electric breakdown using reaxFF forcefields gives a great promise of addressing electron-collisions processes in plasma physics. Numerous plasma applications are now accessible using MD simulations since plasma interactions are of atomic and molecular nature. Moreover, MD simulations allow to identify molecular scale mechanisms that impact real process, for example identification of plasma microstructure of deposited films, reaction pathways in plasma chemistry in various area (plasma medicine, plasma treatment of wastewater, etc). Finally, Molecular dynamics is ultimately one of the links in a multi-scale model of materials in the face of plasma \cite{Bonitz2019}

\backmatter

\bmhead{Acknowledgments}

I wish to warmly thank all people, either experimentalist, theoreticist or numerician, co-authors and/or colleagues, from GREMI, from France and abroad, being staff members, PhD candidates or postdoc,  who contribute by running simulations or experiments and by stimulating discussions to the field of Molecular Dynamics simulations in plasma processing: Anne-Lise Thomman, Ama\"el Caillard, Jean-Marc Bauchire, Johannes Berndt, Eva Kovacevic, Olivier Aubry, Dunpin Hong, Hervé Rabat, Eric Robert, Maxime Mikikian, Lu Xie, Lucile Pentecoste, Soumya Atmane, Mathieu Mougenot, Andrea Jagodar, Sotheara Chuon, Glenn C. Otakantza-Kandjani, Gautier Tetard, Amal Allouch, Matthieu Wolff, Rui Qiu, Fanchao Ye, Jehiel Nteme-Mukunzo, William Chamorro-Coral, Vanessa Orozco-Montes, Sara Ibrahim, Seyedehsara Fazeli,  Christine Charles, Rod W. Boswell, David B. Graves, Erik C. Neyts, Monica Magureanu, Corina Bradu, Magdalena Nistor, Florin Gherendi, Movaffaq Kateb, Tomas Gudmunsson, Sudeep Bhattarcharjee, Christophe Coutanceau, Khaled Hassouni, Armelle Michau, Emilie Despiau Pujo, Tatiana Itina, Germain Valverdu, Marjorie Cavarroc, Pascal Vaudin.

\section*{Declarations}
\begin{itemize}
\item The author declares no conflict of interest
\end{itemize}

\bibliography{RMPP-D-23_00030R1}% common bib file

\end{document}